\documentstyle[twocolumn,aps,floats]{revtex}

\begin{document}

\draft \tolerance = 10000

\setcounter{topnumber}{1}
\renewcommand{\topfraction}{0.9}
\renewcommand{\textfraction}{0.1}
\renewcommand{\floatpagefraction}{0.9}
\newcommand{\br}{{\bf r}}

%Fixing abstract in twocolumn mode
\twocolumn[\hsize\textwidth\columnwidth\hsize\csname
@twocolumnfalse\endcsname

\title{The Multifractal Time and Irreversibility in Dynamic Systems}
\author{L.Ya.Kobelev }
\address{Department of  Physics, Urals State University \\ Lenina
Ave., 51, Ekaterinburg 620083, Russia  \\ E-mail: leonid.kobelev@usu.ru }

\maketitle

\begin{abstract}
The irreversibility of the equations of classical dynamics (the Hamilton
equations and the Liouville equation ) in the space with multifractal time
is demonstrated. The time is given on multifractal sets with fractional
dimensions. The last  depends on densities of Lagrangians in a given time
moment  and in a given point of space. After transition to sets of time
points with the integer dimension the obtained equations transfer in the
known equations of classical dynamics. Production of an entropy is not
equally to zero in space with multifractal time, i.e. the classical
systems in this space are non-closed.
\end{abstract}

\pacs{ 01.30.Tt, 05.45, 64.60.A; 00.89.98.02.90.+p.}
\vspace{1cm}

%Fixing abstract in twocolumn mode

]

\section{Introduction}
The dynamic equations of the physical theories are reversible, it is well
known. The kinetic equations of the statistical theory are irreversible.
The irreversibility  the Boltzmann's statistical theory was in due time
the main reason of non-recognition by Poincare of Boltzmann's statistical
theory. For the irreversibility introducing  in the physical dynamical
equations (for example, in the equation of the Liouville)  it is necessary
to introduce the dissipation terms \cite{Bolch}, \cite{Klim} or the
functionals of a microscopic entropy and time \cite{Prig} ensuring
realization of the second law of thermodynamics. In the first case
irreversibility in the dynamic equations arises as a sequence of
mathematical approaches. Prigogin's the point of view is consists in
recognizing the primacy of irreversible processes and it seems intuitively
more reasonable. Are more general, than mentioned, a methods of
introducing of the irreversibility in the dynamical equations of physics
exist? Is it possible the reversibility of the equations of the dynamical
theories to introduce as result of approximate transition from more
rigorous the dynamical irreversible equations (these dynamic equations may
be obtained as a result of generalization of the known equations) to the
idealized and reversible, but approximate  equations? The purpose of a
note is to introduce one of a possible generalizations of the dynamical
theories of physics realized by replacement of time with topological
dimension  equal unity on "multifractal" time (for the first time it was
introduced in \cite{Kob1}). In the mentioned theory the time is
characterized in each time and space points by  fractional (fractal)
dimensions (FD)$d_{t}(\br(t),t)$. The marked replacement dimension of time
by fractional dimensions gives in the origin of irreversibility in the
dynamical equations of physics and  the existence of irreversibility in
our world may be interpreted by new reason. It is not contradicts (for FD
is small differs from unity for weak physical fields on the Earth) an
experimental data, and allows to receive a new interesting physical
results. The equations of the classical mechanics (the equations of the
Hamilton - Liouville) are chosen for research as an typical  example of
dynamic systems.

\section{Equation of a mechanics with time defined on multifractal sets}

The method of generalization of the classical mechanics equations is
founded on the new model of an approach to the problem of a nature of time
\cite{Kob1}.This model is consists in replacement of usual time by the
time defined on a multifractal subsets $s_{t}$ of continuously set $M_{t}$
(the measure carrier). The multifractal set $ S_{t}$  consists of subsets
$s_{t}$,  i.e. very small time intervals (in further named "points"),
which are also multifractal, and each of them is characterized in turn by
its global fractal dimension {FD) $d_{t}(\br(t),t)$  ( defined as
box-dimension, \cite{Haus}, \cite{Renyi} and so on) that depends  of a
nature of sets $s_{t}$, and depends at coordinates and time (see
\cite{Kob1}, \cite{Kob2}). So each time  subset is characterized by its
global fractal dimension $d_{t}(\br(t),t)$  which characterize the scaling
characteristics for this subset. The continuity $d_{t}(\br(t),t)$ is
supposed. The new approach to a nature of time is consists in the
replacing the usual time points of time axe by selection for describing of
the time's intervals (the "points" on a time axis consisting of sets
$s_{t}$ that defined on the measure carrier $M_{t}$)  only the "points"
that characterized by sets $s_{t}$. The time axe (or, in other selections
time plane or time volume $R_{n}$) is the carrier of measure of all the
multifractal time subsets $s_{t}$ defined on it. The researching of the
dynamic equations and physical quantities (in particular, the entropy)
with time "points" with fractal characteristics defined on multifractal
sets $s_{t}$ with FD $d_{t}(\br(t),t)$, lead to irreversibility of all
dynamic theories used in physics. It is stipulated by an openness of
dynamic systems with multifractal time (role of a thermostat plays set
$R_{n}$) that is appears in a time dependencies of all physical and
mathematical (except for zero) objects. For describing of small changes of
functions defined on multifractal time sets, it is impossible to apply
ordinary or fractional (in sense of the Riemann - Liouville) derivatives
and integrals, since to different time points there corresponds to
different fractal dimensions. For describing of changes of such functions
need's introduction of generalized fractional derivatives and integrals
\cite{Kob1}. In this note the multifractal properties of the space sets
$s_{\br}$ are not considered, since the irreversibility of the dynamic
physical equations arises already at the using only of multifractal time
(see also \cite{Kob2})and global FD of small time sets intervals $s_{t}$.

\section{Generalized fractional derivatives and integrals on multifractal
set $S_{t}$ of time points}

It is necessary if we want to describe the dynamics of time-dependent
functions determined on multifractal set $S_{t}$  to enter the functionals
that extends the fractional derivatives and integrals of the Riemann -
Liouville on the set $S_{t}$ with FD $d_{t}(\br(t),t)$) that is different
in each subsets $s_{t}$ \cite{Kob1} )
\begin{equation}\label{eq1}
D_{+,t}^{d_t}f(t)=\left({\frac{d}{dt}}\right)^n \int\limits_a^t
{dt'\frac{f(t')}{\Gamma(n-d_t(t'))(t-t')^{d_t(t')-n+1}}}
\end{equation}

\begin{eqnarray}\label{eq2} \nonumber
& & D_{-,t}^{d_t}f(t)=(-1)^n \times \\ & & \times
\left({\frac{d}{{dt}}}\right)^n\int\limits_t^b
{dt'\frac{{f(t')}}{{\Gamma(n-d_t(t'))(t'-t)^{d_t(t')-n+1}}}}
\end{eqnarray}

where $\Gamma$-is Euler gamma -function,$a<b$, $a$ and $b$ is stationary
values selected on an axis (from $-\infty$ to $\infty$), $n-1 \leq
d_{t}<n$, $n=\{d_{t}\}+1$, $\{d_{t}\}$ is an integer part of $d_{t}\geq
0$, $n=0$ for $d_{t}<0$, $d_{t}=d_{t}(\br(t),t)$-is the fractal dimensions
(FD). The dependencies  FD from time and space coordinates are defined by
the Lagrangians's densities of a viewed problem \cite{Kob1}, \cite{Kob2},
\cite{Kob3}. The generalized fractional derivative (GFD)
(\ref{eq1})-(\ref{eq2}) coincide with fractional derivatives or fractional
integrals of the Riemann - Liouville \cite{Samko} in the case
$d_{t}=const$. At $d_{t}=n+\varepsilon(t)$, $\varepsilon \to 0$ GFD are
represented by usual derivatives and integrals \cite{Kob1}. The functions
and integrals in (\ref{eq1})-(\ref{eq2}) are considered as generalized
functions given on the set of finitary functions \cite{Gelf}. The
definitions the GFD (\ref{eq1})-(\ref{eq2}) allow to describe the dynamics
of functions defined on multifractal sets and GFD substitute (for such
functions) the  usual or fractional differentiation and integration (GFD
partially conserve the memory about of the last time events).

\section{Hamilton equations}
The Hamilton equations for system from $N$ of classical particles with
identical masses m on the set with multifractal time (i.e. time defined on
multifractal set $S_{t}$) reads:
\begin{equation}\label{eq3}
D_{+,t}^{d_t}\br_i=\frac{\partial H}{\partial {\bf p}_i}, \quad
D_{-,t}^{d_t}{\bf p}_i=-\frac{\partial H}{\partial \br_i}, \quad {\bf
p}_i=D_{+,t}^{d_t}\br _i
\end{equation}
\begin{equation}\label{eq4}
H=\sum\limits_{i=1..N}{\frac{{\bf p}_i^2}{2m}}+\frac{1}{2}\sum\limits_{i
\ne j=1..N}{V(\left|{\br_i-\br_j}\right|})
\end{equation}

The equations (ref{eq3})-(ref{4}) differ from the classical Hamilton
equations by replacement the derivatives with respect to time by GFD
(ref{eq1})-(ref{2}) and coincide with the classical equations of a
mechanics at $d_t=1$. The equations for arbitrary function $B$ dynamic
variable ${\bf p},\br$  will look like
\begin{equation}\label{eq5}
D_{+,t}^{d_t}B=\tilde D_{+,t}^{d_t}B+\frac{\partial H}{\partial{\bf p}_i
}\frac{\partial B}{\partial \br_i}-\frac{\partial H}{\partial \br_i}
\frac{\partial B}{\partial {\bf p}_i }
\end{equation}
The figure $\tilde D_{+,t}^{d_t}$ in (\ref{eq5}) differs from
$D_{+,t}^{d_t}$ in (\ref{eq1}) by replacement the complete derivative with
respect to time $t$ on a partial differential with respect to time $t$.
Let's show, that the Hamiltonian function $H$ in space with multifractal
time is not integral of a motion of the equation (\ref{eq5}), i.e. does
not convert a right part of (\ref{eq5}) in zero. Substitution $H$ in
(\ref{eq5}) gives in
\begin{equation}\label{eq6}
D_{+,t}^{d_t}H=\tilde D_{+,t}^{d_t}H
\end{equation}
From the (\ref{eq1}) follows, that equation (\ref{eq6}) is of the form
(for $d_t(\br (t),t)=1+\varepsilon (\br (t),t)$, $\varepsilon \to 0$, when
a simplifying assumption about lack at $d_t$, $\varepsilon$ of explicit
dependence from $t$ is valid)
\begin{equation}\label{eq7}
D_{+,t}^{d_t}H=\tilde D_{+,t}^{d_t }H=\frac{\varepsilon H}{\Gamma
(1+\varepsilon)t^{d_t}}
\end{equation}
and is equal to zero when $d_t(\br (t),t)=1$. So, in space with
multifractal time, at classical system with a Hamiltonian $H$ that not
depends explicitly at time (conservative systems) the GFD with respect to
a total energy depends on time and decreases with  the increases of time,
i.e. in the model of multifractal time the rigorously conservative
classical systems does not exist. For differs $d_t(\br (t),t)$ from unity
by a little bit (that it is valid about it represents experimental data
about time and results of \cite{Kob1} ) the changing of energy of system
will be very small. Let's consider the problem of change $H$ with change
of time in more general case ($\varepsilon=\varepsilon(\br (t),t)$). For
this purpose we shall be restricted to a case, when FD of  time $d_t(\br
(t),t)=1$ is not considerably differs from unity: $d_t=1+\varepsilon(\br
(t),t)=1$, $|\varepsilon|\ll 1$. In this case GFD  is represented as
\cite{Kob1} (integral is calculated as the total of a principal value and
residue in a singular point)
\begin{eqnarray}\label{eq8} \nonumber
\tilde D_{+,t}^{1+ \varepsilon}H &\approx& \frac{\partial}{\partial t}H\mp
\frac{1}{2}\frac{\partial}{\partial t}[\frac{\varepsilon (\br (t),t)H}
{\Gamma(1+\varepsilon(\br (t),t))}]+ \\ &+& \frac{\varepsilon H}{\Gamma(1+
\varepsilon)t^{d_t}}, \quad\varepsilon>0,\quad d_t<1
\end{eqnarray}
\begin{eqnarray}\label{eq9} \nonumber
\tilde D_{+,t}^{1+\varepsilon}H &\approx& \frac{\partial}{\partial t}
[\frac{1}{\Gamma(1-\varepsilon)}H] \pm \frac{1}{2} \frac{\partial}
{\partial t}[\frac{\varepsilon(\br (t),t)H}{\Gamma(1-\varepsilon)}]+
\\ &+& \frac{ \varepsilon H}{\Gamma(1+\varepsilon)t^{d_t}}, \quad
\varepsilon>0, \quad d_t>1
\end{eqnarray}
The selection of signs (plus or minus) in (\ref{eq8})-(\ref{eq9}) is
determined by sign of $\varepsilon$ and requirements of a regularization
of integrals and selection of FD $d_t$ (greater or smaller unity). Let
$d_t(\br (t),t)<1$. In this case from (\ref{eq8}) follows (for $H$ do not
containing explicit time dependence)
\begin{equation}\label{eq10}
D_{+,t}^{d_t}H=\tilde D_{+,t}^{d_t}H\approx\pm\frac{1}{2}H\frac{\partial
\varepsilon(\br (t),t)}{\partial t}+\frac{\varepsilon H}
{\Gamma(1+\varepsilon )t^{d_t }}
\end{equation}
 For $t \to \infty$, the basic contribution in the (\ref{eq10}) imports
corrections proportional to velocity of change FD $d_t(\br (t),t)$. The
total energy conservative (in sense, that the Hamiltonian has not an
explicit dependence at time ) systems now in space with multifractal time
is not conservative systems. It changes can be at $t \to \infty$ of any
sign, and depend on a sign of derivatives with respect to time from the
fractional correction to dimension of time $\varepsilon$. For
$\varepsilon=0$ the total energy of system is conserves and all relations
coincide with known relations following from the dynamic equations of
classical systems mechanics.

\section{Liouville equation}
The equation of the Liouville for a N-partial distribution function
$\rho(X,t)$ ($X$- are coordinate and impulses of particles of a
$6N$-dimension phase space) is equivalent to the Hamilton equations  for
system from $N$ of classical particles and is invariant in relation to
transformations
\begin{equation}\label{eq11}
\br_i \to \br_i,\quad {\bf p}_i \to -{\bf p}_i,\quad t \to -t,\quad
i=1,2,3,..N
\end{equation}
(equation is reversible). On set $S_t$ of multifractal time complete
derivative $\rho(X,t)$ will reads:
\begin{eqnarray}\label{eq12} \nonumber
& &D_{+,t}^{d_t} \rho(X,t)=\tilde D_{+,t}^{d_t} \rho (X,t)+ \frac{\partial
H}{\partial {\bf p}_i}\frac{\partial \rho (X,t)}{\partial \br_i}- \\ & &
-\frac{\partial H}{\partial \br_i} \frac{\partial \rho (X,t)}{\partial
{\bf p}_i}=\tilde D_{+,t}^{d_t} \rho(X,t)-L \rho(X,t) \\ \nonumber & &
{\bf p}_i=D_{+,t}^{d_t} \br_i
\end{eqnarray}
where $D_{+,t}^{d_t} \rho$ is defined by (\ref{eq1}), $L$-functional of
the Liouville and differs from the functional of the known equation of the
Liouville by replacement of ordinary derivatives with respect to time on
GFD. For a demonstration of the irreversibility of expression (\ref{eq12})
to transformations (\ref{eq11}) we shall mark the following: the
distribution functions $\rho(X,t)$ and $\rho(X_0,t_0)$ viewed in different
moments $t_0$ and $t$ are connected by the relation
\begin{equation}\label{eq13}
\rho(X,t)dX=\rho(X_0,t_0)dX_0
\end{equation}
in which because of change of fractal dimension with a change time $dX
\neq dX_0$. Therefore $\rho(X,t) \neq \rho(X_0,t_0)$ and  $\rho(X,t)$
evolves with a time. Complete derivative  $D_{+,t}^{d_t}\rho(X,t)$, in
particular and in that connection, is not equal to zero. Intuitively it is
clear, that derivative $D_{+,t}^{d_t}\rho(X,t)$ is determined by a
functional from function $\rho(X,t)$ equal to zero at
$d_t=1+\varepsilon=1$. Let's designate this functional describing change
$\rho(X,t)$ owing to interaction with "thermostat", by
$\varphi=\varphi_0(\rho,d_t,t)\varepsilon(\br (t),t)$. The role of the
thermostat plays the set $M_t$ (being the carrier of a measure of a
subsets of time points $s_t$ and belonging to one of spaces $R^n$), as was
already marked. The appealing of the functional $\varphi$ is caused not to
interior processes happening with change of energy inside system, but is
determined by different properties of time sets $s_t$ in different instant
of time (change of dimension of $s_t$ with a time changes). Complete
derivative in this case will be equal to a functional $\varphi$ and
(\ref{eq12}) will reads as the equation
\begin{eqnarray}\label{eq14} \nonumber
\tilde D_{+,t}^{d_t} \rho(X,t) &+& \frac{\partial H}{\partial {\bf p}_i}
\frac{\partial \rho (X,t)}{\partial \br_i}- \frac{\partial H}{\partial
\br_i} \frac{\partial \rho (X,t)}{\partial {\bf p}_i}= \\ &=& \varphi_0
(\rho,d_t,t) \varepsilon(\br(t),t)
\end{eqnarray}
The equation (\ref{eq14}) is analogous of the Liouville equation of the
classical systems with the time defined on multifractal sets. The analog
of a collision integral in a right member (\ref{eq14}) is stipulated by
interaction with the carrier of a measure of multifractal set $M_t$ and is
equal's to zero if sets of time $s_t$ is substitutes by sets with
topological dimension equal to unity.

\section{Production of an entropy}
Let's consider a classical system which is be found in an equilibrium
state at the usual describing of the time.The  production of an entropy in
such system is equal to zero. Let's consider the production of the entropy
$S= \int{\rho(X,t) \ell n \rho(X,t)dX}$ of same classical system defined
on multifractal set of time points $S_t$ (for $d_t=1-\varepsilon<1)$:
\begin{equation}\label{eq15}
D_{+,t}^{d_t}S=\int{D_{+,t}^{d_t}[\rho \ell n \rho]dX}
\end{equation}
Permissible, as well as earlier, that $d_t=1+\varepsilon(\br (t),t)$,
$|\varepsilon \ll 1|$. As $\rho(X,t)$ has  for the equilibrium system at
$d_t \neq 1$  the complete GFD which is non-equal zero, the right member
(\ref{eq15}) is not equal to zero. It means, that equilibrium systems does
not exist in space with multifractal time. Really, as
\begin{eqnarray}\label{eq16} \nonumber
D_{+,t}^{d_t}S=\int{D_{+,t}^{d_t}[\rho \ell n\rho]dX} &\approx& \\
\frac{\varepsilon S}{t^{d_t}} \pm \frac{1}{2} \frac{\partial}{\partial t}
(\varepsilon S)+\frac{\partial S}{\partial t} \ne 0
\end{eqnarray}
that (\ref{eq16}) is an inequality and in a case $\frac{\partial
S}{\partial t}=0$. For $\frac{\partial \varepsilon}{\partial t}=0$ the
production of the entropy is positive. For $\frac{\partial
\varepsilon}{\partial t} \neq 0$, $\frac{\partial S}{\partial t}=0$ the
production of the entropy can have any sign (in particular, the entropy
can be decreasing too). Let's mark in that circumstance, that all new
results for behavior of the entropy are stipulated by the multifractality
of the time and disappear after transition to time with topological
dimension equal to unity.

\section{About connection FD $d_{t}$ with Lagrangians of physical fields}
In the monograph \cite{Kob1} the following approximating connection of
fractional dimension of time $d_t(\br(t),t)$ with a Lagrangian density $L$
of all physical fields in a point $\br(t)$ in an instant $t$ (see also
\cite{Kob2}, \cite{Kob3}) is obtained:
\begin{equation}\label{eq17}
d_t(\br (t),t)=1+\sum \limits_i{\beta_i L_i(\br (t),t)}
\end{equation}
where $\beta_i$ is dimensional numerical factors ensuring a zero dimension
of products $\beta_i L_i$. In \cite{Kob1} it is shown, that for coinciding
the results founded on the (\ref{eq17}) with results of the theory general
relativity (GR) it is necessary (for gravitational forces) to choose
$\beta=\frac{2}{c^2}$ ($c$- speed of light). For correspondence with
results of a quantum mechanics, for electric fields the $\beta_e$ has an
order of magnitude $\beta_e=(2mc^2)^{-1}$ ($m$ -is mass of particle or
body creating the electric charge). The small differences FD from unity is
satisfied by condition
\begin{equation}\label{eq18}
\sum \limits_i {\beta _i}L_i=\varepsilon \ll 1
\end{equation}
The connection $\varepsilon$ with density of Lagrangians adds physical
sense GFD (more in detail about it see \cite{Kob2}) and renders concrete
relations obtained in the previous paragraphs.

\section{Conclusions}
The present note is devoted to the appendix of idea of the multifractal
time offered in \cite{Kob1}  \cite{Kob3}, for researching of the problem
of an irreversibility in large classical systems consisting from an
identical objects. The following results, leading from this paper, on our
sight, are essential:\\ 1. The equations describing behavior of
conservative systems (in usual time), are irreversible in space with
multifractal time;\\ 2. The neglecting by the fractionality of dimension
of time and transition in space with topological dimension of time equal
to unity allows to receive the known reversible equations of classical
dynamics;\\ 3. In space with the multifractal time there are no invariable
objects, since the  GFD with respect to stationary values are not equal to
zero. From the physical point of view it is the reflection of
non-stationary of the Universe with multifractal time. Last statement
corresponds to mathematical exposition of behavior of physical objects and
not contradict the exposition of the Einstein type of  Universe , in
which, in connection with its expansion, there are no invariable
objects;\\ 4. The quantity of the fractional additional to topological
dimension of time member $\varepsilon$  is determined by physical fields
and depends on the density of energy that presents in the given moment in
the given point of space . At the small densities of energy the
corrections are very small. So, for the gravitational fields at distances
more larger that gravitational radius (for example, for FD created  by
mass of the Earth on a surface of Earth) and for electric fields on atomic
distances the value of $\varepsilon$ is equal $\varepsilon \sim 10^{-8}$.
Therefore the multifractal nature of time ($d_t \sim 1+\varepsilon$) does
not contradicts an existing experimental data.

\end{document}